\documentclass[twocolumn,preprintnumbers,amsmath,amssymb,superscriptaddress]{revtex4}

\usepackage{graphicx}
\usepackage[font={small,it}]{caption}
\usepackage{subcaption}
\usepackage{epstopdf}

\usepackage{bm}
\usepackage{bbm}

\usepackage{amsmath}
\usepackage{amsthm}
\usepackage{amssymb}

\usepackage{lscape}
\usepackage{tabularx}
\usepackage{gensymb}

\usepackage{color}
\usepackage[dvipsnames]{xcolor}
\usepackage[normalem]{ulem}

\usepackage{enumitem}
\allowdisplaybreaks[1] 

\newcommand{\bra}[1]{\langle #1|}
\newcommand{\ket}[1]{|#1\rangle}

\newcommand{\ketbra}[2]{|#1\rangle\!\langle#2|}
\newcommand{\id}{\mathbbm{1}}
\newcommand{\expt}[1]{\langle #1\rangle}

\newcommand{\kB}{k_\mathrm{B}}

\newcommand{\inlineheading}[1]{\textbf{{#1---}}}

\begin{document} 

\setlength{\tabcolsep}{1ex}

\title{Guaranteed energy-efficient bit reset in finite time}

\author{Cormac Browne}
\affiliation{Atomic~and~Laser~Physics, Clarendon~Laboratory, University~of~Oxford, Parks~Road, Oxford, OX13PU, United~Kingdom}

\author{Andrew J. P. Garner}
\affiliation{Atomic~and~Laser~Physics, Clarendon~Laboratory, University~of~Oxford, Parks~Road, Oxford, OX13PU, United~Kingdom}

\author{Oscar C. O. Dahlsten}
\affiliation{Atomic~and~Laser~Physics, Clarendon~Laboratory, University~of~Oxford, Parks~Road, Oxford, OX13PU, United~Kingdom}
\affiliation{Center for Quantum Technologies, National University of Singapore, Republic of Singapore}

\author{Vlatko Vedral}
\affiliation{Atomic~and~Laser~Physics, Clarendon~Laboratory, University~of~Oxford, Parks~Road, Oxford, OX13PU, United~Kingdom}
\affiliation{Center for Quantum Technologies, National University of Singapore, Republic of Singapore}

\date{\today}

\begin{abstract}
Landauer's principle states that it costs at least $\kB T\ln2$ of work to reset one bit in the presence of a heat bath at temperature $T$.
The bound of $\kB T\ln2$ is achieved in the unphysical infinite-time limit. 
Here we ask what is possible if one is restricted to finite-time protocols. 
We prove analytically that it is possible to reset a bit with a work cost close to $\kB T\ln2$ in a finite time.
We construct an explicit protocol that achieves this, which involves thermalising and changing the system's Hamiltonian so as to avoid quantum coherences.
Using concepts and techniques pertaining to single-shot statistical mechanics, we furthermore prove that the heat dissipated is exponentially close to the minimal amount possible not just on average, but guaranteed with high confidence in every run. 
Moreover we exploit the protocol to design a quantum heat engine that works near the Carnot efficiency in finite time.
\end{abstract}

\maketitle

\inlineheading{Introduction.}
Landauer's principle~\cite{Szilard29,Landauer61,Bennett82} states that resetting a bit or qubit in the presence of a heat bath at temperature T costs at least kTln2 of work, which is dissipated as heat. 
It represents the fundamental limit to heat generation in (irreversible) computers, which is extrapolated to be reached around 2035~\cite{Frank05}.

The principle is also a focal point of discussions concerning how thermodynamics of quantum and nano-scale systems should be formulated. Of particular interest to us here is the single-shot approach to statistical mechanics~\cite{delRioARDV11,Aberg11,DahlstenRRV11,EgloffDRV12}. 
This concerns statements regarding what is guaranteed to happen or not in any single run of an experiment, as opposed to what happens on average. 
This distinction is important for example in nano-scale computer components, in which large heat dissipations in individual runs of the protocol could cause thermal damage, even if the average dissipation is moderate. 

In~\cite{delRioARDV11} Landauer's principle was assumed to hold in the strict sense that one can reset a uniformly random qubit at the exact work cost of $kT\ln2$ each run of an experiment. 
This assumption can be showed to be justified if one allows quasistatic protocols~\cite{Aberg11,EgloffDRV12}.

Real experiments take place in finite time~\cite{Jarzynski97, SekimotoS97, SerreliLKL07,ThornSLS08,Toyabe2010,AurellMM11, BerutAPCDL12,BergliGK13,Zulkowski13,Koski14}. 
Our present Letter is motivated by the concern that fluctuations might be much greater in finite time scenarios, and that the single-shot optimality expressions to date may therefore not be physically relevant. 
Therefore we extend the protocol for bit reset used in~\cite{delRioARDV11} to the finite time regime and analyse what changes.
In this regime thermalisation is imperfect, and the quantum adiabatic theorem fails so that one cannot a priori assume that shifting energy levels does not change occupation probabilities.
Moreover there are correlations between occupation probabilities at different times. 
We prove analytically that it is in fact possible to reset a qubit in finite time at a guaranteed work cost of $kT\ln2$, up to some small errors.
We also make a natural extension of our results to a qubit heat engine which operates at the Carnot efficiency in finite time up to a small error. 
We derive bounds on the errors, proving that they fall off exponentially or doubly exponentially in the time taken for the~protocol. 

\begin{figure}[h!]
\begin{centering}
\includegraphics[width=0.45\textwidth]{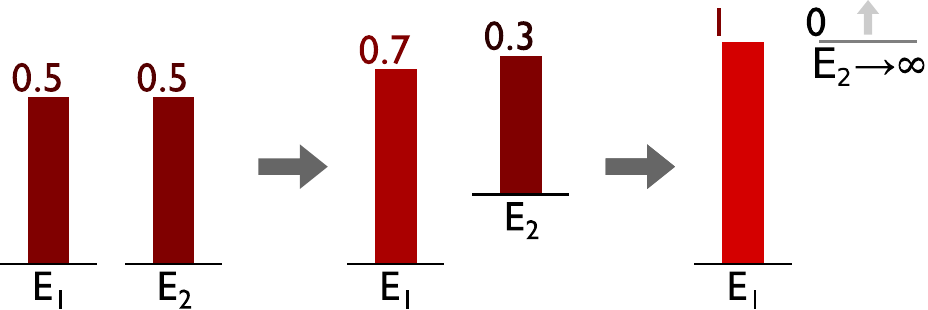}
\caption{
\label{fig:2level}
{\bf `Bit reset' by raising $\mathbf{E_2}$ to infinity.}
The numbers indicate the occupation probability for the energy levels. 
We consider the extension of this protocol, \newline
 running in finite~time.
}
\end{centering}
\end{figure}

\inlineheading{A quasistatic bit reset protocol.} 
We examine a simple two-level system, with access to a heat bath and a work reservoir, which we will manipulate with a time-varying Hamiltonian in the regime as set out by~\cite{AlickiHHH04}.
The evolution of the system takes place through two mechanisms~\cite{AlickiHHH04} (see also~\cite{Kieu04, EgloffDRV12}):
\begin{enumerate}
\item \label{mech:el_change} Changes to the energy spectrum -- identified with work cost/production $\mathrm{d}E_i$ for occupied energy level~$i$, which has no effect on the occupation \mbox{probabilities}.
\item \label{mech:therm} Changes to probability distributions via interactions with a heat bath (thermalisation), with no changes to the energy spectrum, and thus no associated work cost/production.
\end{enumerate}

Initially, the two degenerate energy levels of a random qubit are equally likely to be populated. 
The system is coupled to a heat bath at temperature $T$, and one energy level is quasistatically and isothermally raised to infinity, until the lower energy level is definitely populated (see Fig.~\ref{fig:2level}). 
If we want to perform bit reset (``erasure'' in Landauer's terminology),
 the system is then decoupled from the heat bath, and the second energy level is returned to its original value, 
 such that the initial and final energy level configurations are the same, and only the populations have changed.
In the quasistatic limit, each stage of raising the energy level has an average cost of $P_2 \mathrm{d}E$ where $P_2$ is the (thermal) population of the upper energy level. 
Thus raising the second level from 0 to infinity, the work cost of the entire protocol is given:
\begin{equation}
\label{eq:quasistatic_work}
\expt{W} = \int_0^\infty P_2(E)\mathrm{d}E = \kB T \ln2.
\end{equation}

 
\begin{figure}[th]
\begin{centering}
\includegraphics[width=0.45\textwidth]{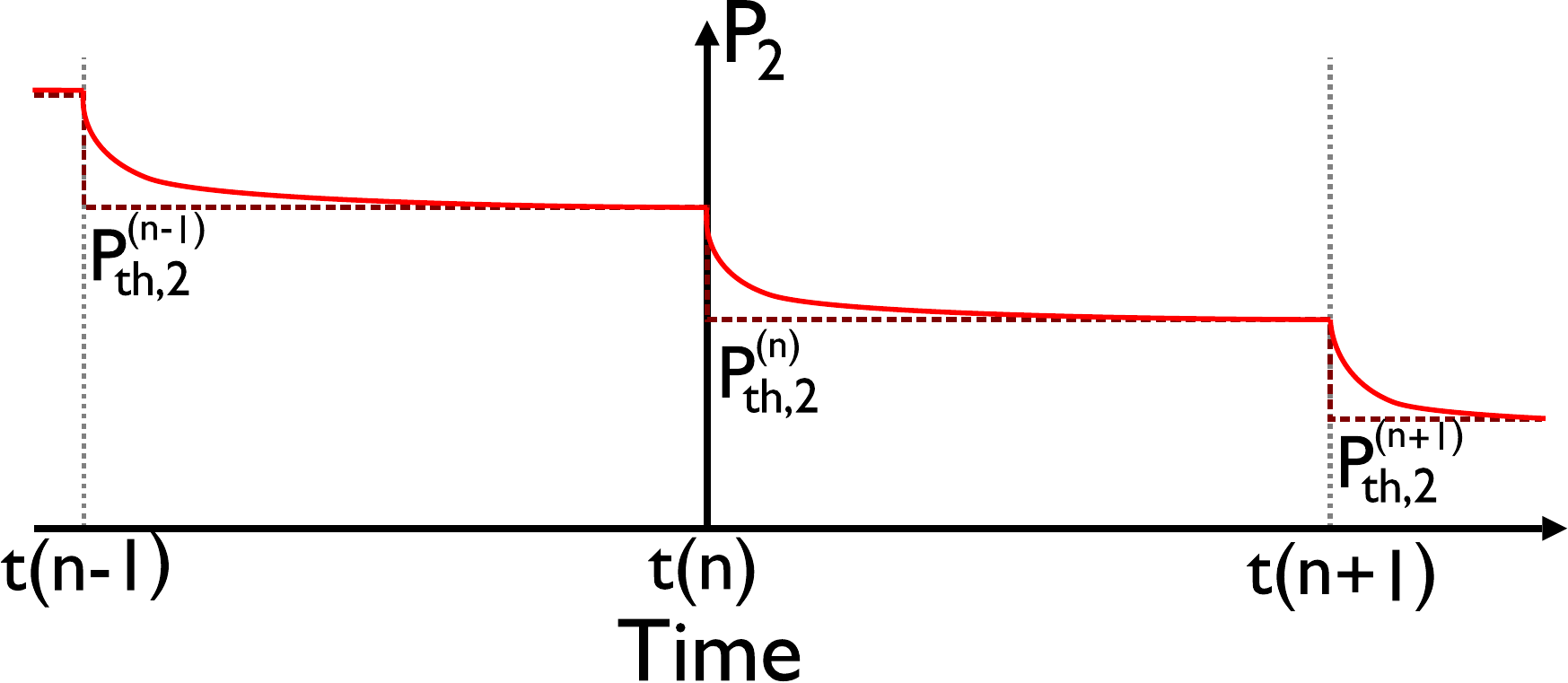}
\caption{
\label{fig:therm_pop}
{\bf A few steps of the protocol.} Thermal (dashed) and actual (solid) upper energy level populations with respect to time. 
At each time, $t(n)$, the upper energy level is raised,  altering the thermal state.
}
\end{centering}
\end{figure}

We now describe how these two elementary steps are impacted by the extension to finite time. 

\inlineheading{Level-shifts in finite time: the negative role of coherence.}
A development of coherences during the level shifts would increase the amount of work that must be invested to perform a reset. 
For example, suppose the initial state is $\rho_i=p(a)\ket{a}\bra{a}+ p(b)\ket{b}\bra{b}$, where $\ket{a}$, $\ket{b}$ are the energy eigenstates with occupation probabilities $p(a)$, $p(b)$ respectively. 
Suppose the Hamiltonian changes very quickly and the new energy eigenstates are $\ket{a'}$, $\ket{b'}$. 
Using the scheme of~\cite{quan08}, we perform a projective measurement in the new energy eigenbasis, $\ket{a'}$, $\ket{b'}$ and define the work input of the step as the energy difference between the initial energy eigenstate and the final. 
The average work cost can be written as
\begin{eqnarray}
\label{coherent_work}
\expt{W}\!\! &=& p(b)(p(b \to b')(E_{b}' - E_{b}) + p(b \to a')(-E_b)) \nonumber \\
&& +\hspace{0.5em} p(a)(p(a \to a')(0) + p(a \to b')(E_{b'})) \nonumber \\
&=& p(b)(E_{b}' - E_{b})  + (p(a) - p(b))p(a \to b')E_{b'},
\end{eqnarray}
where $E_i$ is the energy of associated with state $i$, and $p(a \to b')$ indicates the probability of transitioning from $a$ to $b'$. 
(In our protocol, $E_a\!\!=\!\!E_{a'}\!\!=\!0$; the second line follows by noting that the transition probabilities are doubly-stochastic due to the Born rule). 
Since $E_{b'}\!>\!E_b\!\geq\!0$, and $p(a)\!\geq\!p(b)$, in order for this expression to contribute the least work cost possible, we set the transition probability $p(a \rightarrow b')$ to zero, i.e.\ by not causing any \mbox{coherent~excitations}.

We therefore avoid coherence, using either one of two ways: (i) By allowing for the experimenter to choose a path of Hamiltonians that share the same energy eigenstates and only differ in energy eigenvalues, as in the standard model for Zeeman splitting (Appendix~\ref{app:avoidquantum}), (ii) If the experimenter is instead given a fixed Hamiltonian path which is not of this kind, she may actively remove the coherences by an extra unitary being applied (Appendix~\ref{app:reversequantum}). 
This method of active correction is similar to the strategy employed in super-adiabatic processes~\cite{Berry09,Deffner14,DelCampo13}.
In either case, at any point of the protocol the density matrix of the system will be diagonal in the instantaneous energy eigenbasis. 

\inlineheading{Partial thermalisation can be represented by partial swap matrices.}
The stochastic `partial swap' matrix is equivalent to all other classical models of thermalisation for a two level system~\cite{ZimanSBHSG01,ScaraniZSGB02}.
In this model, the heat bath is a large ensemble of thermal states which have some probability $p$ of being swapped with the system state in a given time interval.

We model a period of extended thermalization as being composed of a series of $t$ partial swap matrices, each representing a unit time step. Multiplying these together gives a single swap matrix with probability $P_\mathrm{sw}(t)$ of swapping with the thermal state:
\begin{equation}
\label{eq:swap_prob_stacking}
P_\mathrm{sw}(t) = 1-(1-p)^t.
\end{equation}
(See Appendix~\ref{app:partialswap} for details.) 
The influence of a finite thermalisation time, $t$, manifests as a degradation of the thermalisation quality through a lowered $P_\mathrm{sw}(t)$.

\inlineheading{The average extra work cost of finite time is exponentially suppressed.}
It is possible to quantify to what extent the state at the end of a period of thermalisation differs from a thermal state by employing the trace-distance $\delta(\rho,\sigma) := \frac{1}{2} \mathrm{Tr} ( \sqrt{ ( \rho - \sigma)^{\dagger}(\rho - \sigma )} ) $ (see e.g.~\cite{LevinPE09}). Throughout the protocol we monotonically raise the energy of the second level and thus monotonically decrease the occupation probability of the second level. This makes it possible to bound the trace-distance after a period of thermalisation by the trace-distance between the desired thermal state and the occupation probability with degenerate energy levels:
\begin{equation}
\label{eq:trace_distance_bound}
\delta \leq \left(\dfrac{1}{1+\exp \left( -\beta E_2 \right)}- \dfrac{1}{2} \right) \left( 1 - p \right)^t.
\end{equation}
(Full derivation in Appendix~\ref{app:deltabound}.)

We thus see that the occupation probabilities get exponentially close to to the thermal distribution as a function of the number of time steps spent thermalizing.
It follows that the average work $\expt{W}$ required to raise the upper level from zero to some maximum energy $E_\mathrm{max}$, only thermalising for $t$ time steps at each stage, is bounded~by

\begin{eqnarray}
\label{eq:work_cost}
\expt{W_\mathrm{quasi}}  & \leq \,\expt{W}\,  \leq &
\expt{W_\mathrm{quasi}} + \nonumber \\
&& (1-p)^t \left(\dfrac{E_\mathrm{max}}{2} - \expt{W_\mathrm{quasi}}\right),
\end{eqnarray}
where $\expt{W_\mathrm{quasi}}$ is the quasistatic work cost of raising the second level to $E_\mathrm{max}$: 
\begin{equation}
\expt{W_\mathrm{quasi}} = \kB T\ln\left(\dfrac{2}{1+\exp\left( -\frac{E_\mathrm{max}}{\kB  T} \right) }\right).
\end{equation}
(Full derivation in Appendix~\ref{app:workbound}.)

Eq.~\ref{eq:work_cost} shows that the average extra work cost of reset decreases exponentially in the thermalisation time, $t$.

\inlineheading{Work fluctuations in a single shot reset are doubly exponentially suppressed.}
The preceding discussion applies to the {\em average work cost} over many bit resets. 
However, the average work cost is not always the most useful parameter if we wish to design a physical device which performs a bit reset.
Another practical parameter is the guaranteed work from single shot thermodynamics, which provides an upper bound on the work required per reset, $W_\mathrm{max}$.
One could envision scenarios where if this value is exceeded the device breaks:
either because the work must ultimately be dissipated as heat which above a certain level could damage the device,
or if the work reservoir itself has a finite capacity and would be exhausted by an attempt to draw on too large a value.

To derive such a parameter, or other similar parameters (including the average work value), it is useful to generate a {\em work distribution}: the probability density function for the work cost of the bit reset process.

It has been established in the quasistatic case that the work cost in any procedure involving shifting energy levels (such as bit reset) is tightly peaked around the average value~\cite{Aberg11,EgloffDRV12}. 
Applying the method of Egloff and co-authors~\cite{EgloffDRV12}, the deviation from the average work in the quasistatic case
when raising the second level by $\mathcal{E}$ per step over $N$ steps is bounded by:
\begin{equation}
\label{eq:ss_work_quasi}
P(|W-\expt{W}| \geq \omega) \hspace{0.5em} \leq \hspace{0.5em} 2\exp\left( - \dfrac{2 \, \omega^2}{N \mathcal{E}^2} \right).
\end{equation}

This result arises from modelling the work input as a stochastic process, and applying the McDiarmid inequality~\cite{McDiarmid89} which bounds the deviation from the mean value.
By thermalising fully in every step, the question of which energy level is occupied at each step can be represented by a sequence of independent random variables with distributions given by the thermal population associated with the particular energy gaps. 
The work cost is a function of these random variables.

We note that altering the energy level at one stage in the perfectly thermal regime will only affect the total work cost by the difference associated with that one step.
However with partial thermalisation, the states the system is in at two different stages are not independent of each other. 
Altering the energy level of one stage has a knock-on effect on the work cost for all subsequent stages. 
We show in Appendix~\ref{app:fluxbound} that this effect falls off in a manner inversely proportional to the probability of swapping,
and thus derive a new finite time version of Eq.~\ref{eq:ss_work_quasi}, which explicitly takes into account this sensitivity:
\begin{equation}
\label{eq:ss_work_partial}
P(|W-\expt{W}| \geq \omega) \hspace{0.5em} \leq \hspace{0.5em} 2\exp\left( - \dfrac{2 \,  \omega^2{P_\mathrm{sw}(t)}^2}{N \mathcal{E}^2} \right).
\end{equation}
When $t\to\infty$ such that $P_\mathrm{sw}(t)\to1$, we recover the perfect thermalisation case. 
As $P_\mathrm{sw}(t)$ is itself an exponential function of time (Eq.~\ref{eq:swap_prob_stacking}), we see this limit rapidly converges on the quasistatic variant as $t$ increases. Hence, fluctuations are supressed as a double exponential function of time spent thermalising.
When $P_\mathrm{sw}=0$ the bound becomes meaningless, as changing what happens at one stage in the protocol can have an unbounded effect on the final work cost.

We can combine this spread with the average result in Eq.~\ref{eq:work_cost} to determine $W^\epsilon_\mathrm{max}$: the maximum work cost in a single shot, guaranteed only to be exceeded with failure probability $\epsilon$ (mathematically, $W^\epsilon_\mathrm{max}$ is defined to satisfy $\int_{-\infty}^{W^\epsilon_\mathrm{max}} P(W) \mathrm{d}W = 1-\epsilon$ for work distribution $P(W)$):
\begin{eqnarray}
\label{eq:guaranteed_work}
W_\mathrm{max}^\epsilon & \leq & \left(1-(1-p)^t\right)\expt{W_\mathrm{quasi}} +\dfrac{1}{2}(1-p)^t E_\mathrm{max} \nonumber \\
&&+ \hspace{1em}\dfrac{1}{1-(1-p)^t} \sqrt{\dfrac{\ln \left( 2/\epsilon \right)}{2N}} E_\mathrm{max}. \label{eq:w_max}
\end{eqnarray}
(Full derivation in Appendix~\ref{app:wemax}.)
From Eqs.~\ref{eq:work_cost},~\ref{eq:ss_work_quasi} \& ~\ref{eq:guaranteed_work} we see that spending time thermalising has two-fold benefit: it lowers the average cost, and reduces the spread of possible work values around this average.


\inlineheading{Reset error is exponentially suppressed.}
Initially, we have complete uncertainty about the system - the energy levels are degenerate and the particle is equally likely to occupy either of them.
At the end of the raising procedure, the system has been `reset' to one energy level, with probability of failure, $P_{fail}$ (defined as the occupation probability of the second energy level), upper bounded by (using the probability distribution derived in Appendix ~\ref{app:deltabound})
\begin{equation}
P_\mathrm{fail} \leq \left(1 + (1-p)^t \right) \dfrac{\exp \left(- \beta E_{max} \right)}{1 + \exp \left(- \beta E_{max} \right) } - \dfrac{1}{2}(1-p)^t.
\end{equation}

\inlineheading{The work cost of resetting multiple bits scales favourably.} 
In \cite{delRioARDV11} the general protocol involves resetting $n$ qubits in the state $\rho=\id/2\otimes\id/2\cdot\cdot\cdot \otimes \id/2$ and we now consider how the errors scale in this case.

Recall that for a single qubit, thermalising for time $t$ bounds the trace-distance $\delta$ between the actual state and the true thermal state according to Eq.~\ref{eq:trace_distance_bound}. 
This results in an additional work cost of up to $\delta E_\mathrm{max}$ across the protocol. For $n$ bits, in the very worst case each bit will be $\delta$ away from its thermal state, and so trivially the extra work cost will be bounded by $n \delta E_\mathrm{max}$.
Thus the average work cost of reset scales linearly with the number of bits (as with the quasistatic case). This argument can be naturally modified to the single-shot case (more detail in Appendix~\ref{app:manyqubits}), and one may therefeore say that the the multi-qubit statements in~\cite{DahlstenRRV11,delRioARDV11} also remain relevant in the case of finite time.

\inlineheading{Work extractable in finite time.}
Although we have thus far concentrated on bit reset, it is also possible to consider the inverse protocol: work extraction by lowering the upper energy level in contact with a hot bath. 
The derivation of this proceeds exactly as before, with the limits of the integration reversed. 
This allows us to extract work bounded by: 
\begin{eqnarray}
\label{eq:work_extraction}
-\expt{W_\mathrm{quasi}} &  \leq & \expt{W_{out}} \nonumber \\ 
& \leq & -\expt{W_\mathrm{quasi}} + \nonumber \\
&& \hspace{0.5em} (1-p)^t \left(\dfrac{E_\mathrm{max}}{2} - \expt{W_\mathrm{quasi}}\right).
\end{eqnarray}

\begin{figure}[t!]
\begin{centering}
\includegraphics[width=0.45\textwidth]{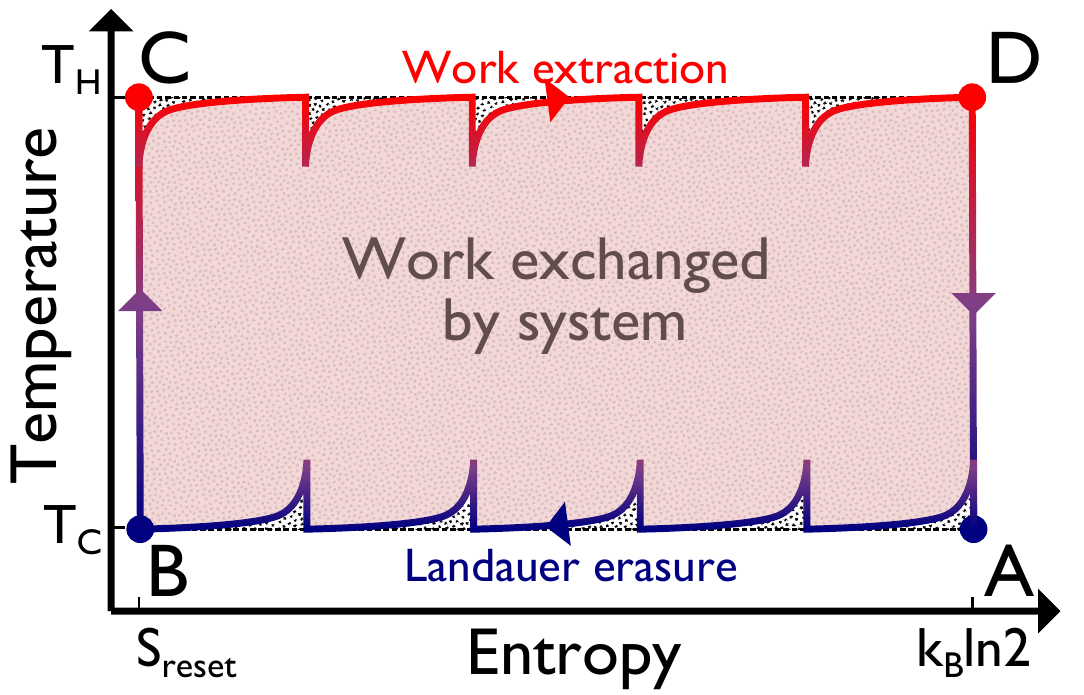}
\caption{
\label{fig:TS_cycle}
{\bf Entropy-temperature diagram for the engine~cycle.}
The block shaded area shows the net work exchanged by the system, which is less than that of the quasistatic (Carnot) cycle shown by the dashed rectangle filled with dots behind.
For illustrative purposes, the number of energy level adjustments has been greatly reduced, and size of the associated shifts in temperature greatly exaggerated.
}
\end{centering}
\end{figure}

If we have access to two baths at different temperatures, we can extract net work 
by running a bit reset coupled to a cold bath at temperature $T_{C}$ followed by a work extraction coupled to a hot bath at temperature $T_{H}$. 
The engine cycle formed (shown in Fig.~\ref{fig:TS_cycle}) has net work exchange per cycle bounded~by
\begin{eqnarray}
\expt{W_\mathrm{net}} & \leq  &-P_{sw}(t)\kB\left(T_{H}-T_{C}\right)\ln 2 \nonumber \\ 
&& -P_{sw}(t)\kB \left(T_{C}\ln\left(Z_{C}^\mathrm{max}\right) + T_{H}\ln\left(Z_{H}^\mathrm{max}\right)\right)\nonumber \\
&& + \left(1-P_{sw}(t)\right) E_\mathrm{max}, \nonumber \\
& & \hspace{-1em}= P_{sw}(t)\expt{W_\mathrm{quasi}} + \left(1-P_{sw}(t)\right) E_\mathrm{max},
\end{eqnarray}
where  $Z^{max}_{C/H}=1+\exp\left( -E_\mathrm{max}/\kB T_{C/H}\right)$  
and $\expt{W_{net}}$ is the difference between the work input (Eq.~\ref{eq:work_cost}) and the work output (Eq.~\ref{eq:work_extraction}) with the appropriate temperatures.

The final line follows because the first two lines are the same as in the quasistatic case.
The final term is independent of the temperatures of the baths; but instead accounts for the effect of finite time on the degree of over-population of the upper level when raising, and under-population of the upper level when lowering.

Adjusting the energy level in finite steps has necessitated a move away from the truly isothermal process of the quasistatic regime.
In a two-level system, any distribution where the lower energy level is more populated than the upper may be written as a thermal state for some temperature.
During bit reset, after the upper energy level is raised, it is over-populated with respect to the thermal population associated with the cold bath temperature.
We can interpret this as a rise in the system temperature appearing as the saw-tooth pattern in~Fig.~\ref{fig:TS_cycle}.

\inlineheading{There is a speed limit for positive power output.}
By reconciling the concept of time with these processes, we can now talk meaningfully about the {\em power} of the cycle, $\mathcal{P}$: the net work exchange divided by the total time taken to complete a full cycle of bit reset and extraction.
The total time is found by multiplying the amount of time spent thermalising in one stage, $t$, by the number of stages in both halves of the protocol, $N = 2 E_\mathrm{max}/\mathcal{E}$.
We can upper bound the power of the entire cycle as a function of the maximum energy gap $E_\mathrm{max}$ and of $t$: 
\begin{equation}
\label{eq:finite_power_short}
\mathcal{P} \leq \dfrac{P_{sw}(t)\expt{W_\mathrm{quasi}} + \left(1-P_{sw}(t)\right) E_\mathrm{max}}{2 \dfrac{E_\mathrm{max}}{\mathcal{E}} t}.
\end{equation}

A positive $\mathcal{P}$ indicates the system will draw energy in from its surroundings, and therefore we note that the engine does not produce work in all parameter regimes. 
If one attempts to operates an engine (with $\expt{W_\mathrm{quasi}} < 0$) quicker than the limit
\begin{equation}
\label{eq:neg_work}
t < \dfrac{-1}{\ln(1-p)}\ln\left(1-\dfrac{E_\mathrm{max}}{\expt{W_\mathrm{quasi}^\mathrm{net}}}\right),
\end{equation}
then the partial thermalisation can potentially contribute an excess work cost greater than the engine's quasistatic work output.

\inlineheading{Near-Carnot efficiency can be achieved in finite time.}
The efficiency $\eta$ of a cycle may be defined as the net work extracted divided by the maximum work value of one bit of information ($-\kB T_{H}\ln2$). 
We write this as
\begin{eqnarray}
\eta_\mathrm{quasi} -  \dfrac{(1-p)^t E_\mathrm{max}}{\kB T_{H}\ln2}
& \leq \eta \leq & \eta_\mathrm{quasi},
\end{eqnarray}
where $\eta_\mathrm{quasi}$ is the quasistatic efficiency of raising to $E_\mathrm{max}$ over an infinitely long time, and can be related to
the Carnot efficiency $\eta_\mathrm{C}$ by
\begin{equation}
\eta_\mathrm{quasi} = \eta_\mathrm{C}
- \dfrac{\ln\left(Z_{H}^\mathrm{max}\right) - \dfrac{T_{C}}{T_{H}}\ln\left(Z_{C}^\mathrm{max}\right)}{\ln2}.
\end{equation}

We see that if $t$ is small, the cost of raising the populated upper energy level takes its toll on the efficiency, potentially plunging it into negative values (work loss) when $t$ does not satisfy Eq.~\ref{eq:neg_work}. 
Conversely, as $t\to\infty$, the process is maximally efficient, but has no~power.

\inlineheading{Conclusion.}
We have showed that one can reset a qubit in finite time at a guaranteed work cost of kTln2 up to some errors that fall off dramatically in the time taken for the protocol. As mentioned, several key results in single-shot statistical mechanics assume that this can be achieved perfectly~\cite{DahlstenRRV11,delRioARDV11,EgloffDRV12,Aberg11,FaistDOR12}. Our Letter accordingly shows that the optimality statements of those papers are still relevant to finite time protocols. Moreover our results naturally extended to show that the Carnot efficiency can be achieved by a qubit engine in finite time up to an error that falls of exponentially in the cycle time.

\inlineheading{Acknowledgments.}
We acknowledge fruitful discussions with Alexia Auff\`eves, Felix~Binder, Dario~Egloff, John~Goold, Kavan~Modi, Felix~Pollock and Nicole~Yunger~Halpern.
We are grateful for funding from the EPSRC (UK), the Templeton Foundation, the Leverhulme Trust, the EU Collaborative Project TherMiQ (Grant Agreement 618074), the Oxford Martin School, the National Research Foundation (Singapore), and the Ministry of Education (Singapore).

\bibliographystyle{h-physrev} 

\clearpage
\newpage
\appendix

\section{Quantum coherences}
\label{app:quantum}
\subsection{Quantum coherence during energy level steps can be avoided.}
\label{app:avoidquantum}

We can avoid inducing quantum coherences when raising energy levels, by choosing to describe the energy levels by Hamiltonians which are diagonal in the same basis at each stage, written as:
\begin{equation}
\mathcal{H}(n) = \left(\begin{array}{cc}
E_1(n) & 0 \\
0 & E_2(n)
\end{array}\right) = \left(\begin{array}{cc}
0 & 0 \\
0 & \mathcal{E} n
\end{array}\right).
\end{equation}
Because they are all diagonal in the same basis, they will always commute; $[\mathcal{H}(n),\mathcal{H}(m)] = 0$ for all $n$ and $m$.
This is a reasonable regime to physically implement, for example, by varying two hyperfine levels in an atom by the application of a time-dependent external magnetic field.

Defining $P_1(n)$ and $P_2(n)$ as the populations at stage $n$ of lower and upper energy levels respectively, we write the density matrix $\rho(n)$ in the energy eigenbasis as:
\begin{equation}
\label{eq:density_rep}
\rho(n) = \left(\begin{array}{cc}
P_1(n) & 0 \\
0 & P_2(n)
\end{array}\right).
\end{equation}

In order to apply the Schr\"odinger equation, we can interpolate into a continuous case and consider the transition from stage $n$ to stage $n+1$ as given by a time parameter $t$ that varies from $0$ to $\tau$ in each step: 
\begin{equation}
\label{eq:energy_step_H}
\mathcal{H}(n, t) = \left(\begin{array}{cc}
0 & 0 \\
0 & \mathcal{E} \left(n + t/\tau \right)
\end{array}\right),
\end{equation}
such that $\mathcal{H}(n, \tau) = \mathcal{H}(n+1, 0)$.

By using the Schr\"odinger equation to generate an equation for unitary evolution from stage $n$ to stage $n+1$, as we have chosen a $\mathcal{H}$ which commutes with itself throughout the process, we can write down:
\begin{equation}
U(n\to n\!+\!1) = \exp \left( -\dfrac{i}{\hbar} \int_0^\tau \mathcal{H}(t) \mathrm{d}t \right).
\end{equation}
$\mathcal{H}(t)$ is always diagonal 
and so it follows that $\int_0^\tau \mathcal{H}(t) \mathrm{d}t$ must also be diagonal. 
Taking the exponent of this sum multiplied by a constant (${i}/{\hbar}$) implies that $U$ must also be a diagonal operator.

The evolution of a mixed state from time $t_i=0$ (i.e. step $n$) to time $t_f$ can thus be written as
\begin{equation}
\label{eq:unitary_ev}
\rho(n\!+\!1) = U(n\to n\!+\!1) ~ \rho(n) ~ U(n\to n\!+\!1)^\dag.
\end{equation}
Providing we always start in a diagonal state, all terms on the right hand side are diagonal, and thus $\rho(m)$ is diagonal for all $m$.

In fact, we can go further and explicitly calculate $U$ due to the simple form of Eq.~\ref{eq:energy_step_H}:
\begin{eqnarray}
U(n\to n\!+\!1) & = & \exp \left( -\dfrac{i}{\hbar} \left(\begin{array}{cc}
0 & 0 \\
0 & \mathcal{E} \left(n+\frac{1}{2}\right)\tau
\end{array}\right) \right), \nonumber\\
& = & \left( \begin{array}{cc}
1 & 0 \\
0 & \exp\left( -\frac{i}{\hbar} \mathcal{E} \left(n+\frac{1}{2}\right)\tau \right)
\end{array}\right).
\end{eqnarray}
This shows us that the only effect of evolution is to introduce a phase between the upper and lower energy levels, $\phi = \frac{\mathcal{E}}{\hbar} \left(n+\frac{1}{2}\right)\tau$.
If the upper and lower energy levels are an incoherent mixture, then this transformation will have no effect whatsoever on the final state; rewriting Eq.~\ref{eq:density_rep} as $\rho(n) = P_1(n) \ketbra{E_1(n)}{E_1(n)} + P_2(n) \ketbra{E_2(n)}{E_2(n)}$, the evolution in Eq.~\ref{eq:unitary_ev} reduces to
\begin{eqnarray}
\rho(n\!+\!1) \hspace{-2.5em} & & \nonumber \\
& = & U\left( P_1(n) \ketbra{E_1(n)}{E_1(n)} + P_2(n) \ketbra{E_2(n)}{E_2(n)} \right) U^\dag, \nonumber \\
& =& P_1(n) \ketbra{E_1(n)}{E_1(n)} + P_2(n)  e^{-i\phi} \ketbra{E_2(n)}{E_2(n)} e^{i\phi}, \nonumber \\
& =& P_1(n) \ketbra{E_1(n)}{E_1(n)} + P_2(n) \ketbra{E_2(n)}{E_2(n)}, \nonumber \\
& =& \rho(n).
\end{eqnarray}

\subsection{The cost of coherent excitations can be actively reversed.}
\label{app:reversequantum}
If the different Hamiltonians are not intrinsically diagonal, we can consider an energy level step as taking us from the initial Hamiltonian $H = \sum_i E_i \ketbra{E_i}{E_i}$ to the final Hamiltonian $\tilde{H} = \sum_i \tilde{E_i} \ketbra{\tilde{E_i}}{\tilde{E_i}}$, where not only have the eigenvalues changed, but there is also a change of basis.
A change in basis can always be associated with unitary $U$ such that
\begin{equation}
\ketbra{\tilde{E_i}}{\tilde{E_i}} = U\ketbra{E_i}{E_i} U^\dag,
\end{equation}
and one can express $\tilde{H} = \sum_i \tilde{E_i} U \ketbra{E_i}{E_i} U^\dag$.

\begin{figure}[htb]
\begin{centering}
\includegraphics[width=0.45\textwidth]{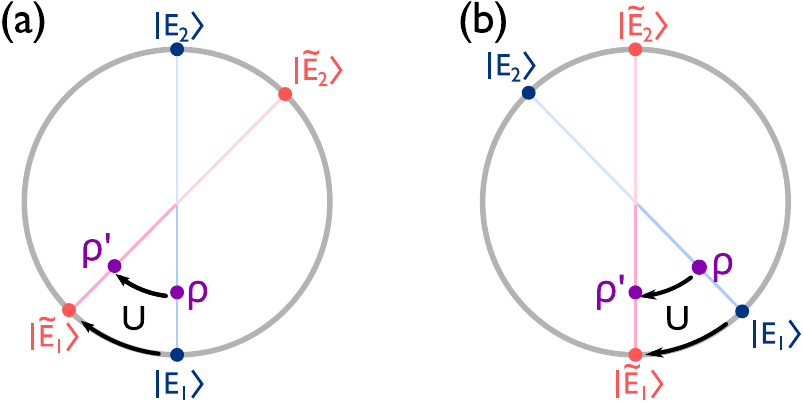}
\caption{
\label{fig:coherent_excitation}
{\bf Coherent excitation.}
(a) shows the state in the Bloch sphere of the initial Hamiltonian, and (b) of the final.
The state $\rho$, diagonal in the initial Hamiltonian, is a {\em coherently excited} state in the final Hamiltonian.
The same passive transformation $U$ between initial and final Hamiltonians can be applied actively on $\rho$ to map it to $\rho'$, compensating for the work cost of coherent excitation.
}
\end{centering}
\end{figure}

Consider the initial state of the system, given by a density matrix written diagonally in the initial Hamiltonian basis: $\rho = \sum_i P_i \ketbra{E_i}{E_i}$.
The system energy before the change in Hamiltonian is defined as $\mathrm{Tr}\left(\rho H \right)$ and afterwards as $\mathrm{Tr}\left(\rho \tilde{H} \right)$, and so we see that the change in system energy (and thus by conservation of energy, the work cost from reservoir) is given by
\begin{equation}
\Delta W = \mathrm{Tr}(\rho \tilde{H} ) - \mathrm{Tr}(\rho H ).
\end{equation}
A visualisation that this might result in an work cost is apparent in Figure~\ref{fig:coherent_excitation} that the state $\rho$ appears higher in the sphere in terms of $\tilde{H}$ than $H$. 
One must take care, however, to note that the energy scale along the vertical axis of each Bloch sphere is different. 
For bit reset, the energy gap is increasing and $\tilde{E_2} - \tilde{E_1} > E_2 - E_1$.

If we were to let the system decohere, the final work cost would be exactly this value (the density matrix in the first term is replaced by a new density matrix with the terms off-diagonal in $\tilde{H}$ removed; but as these terms only appear inside a trace, they do not contribute to the value).

However, if instead we apply an active unitary transformation on $\rho$ to bring it into a new state $\rho'$, which is diagonal in the new energy eigenbasis
$\rho = \sum_i P_i \ketbra{E_i}{E_i} =  \sum_i P_i U^\dag \ketbra{\tilde{E_i}}{\tilde{E_i}} U$ and so the obvious choice is to take $\rho' = U\rho U^\dag$ given as
\begin{equation}
\rho' = \sum_i P_i \ketbra{\tilde{E_i}}{\tilde{E_i}}.
\end{equation}

If we now consider the overall change in energy between the initial state before the change in Hamiltonian, and the final state after the change in Hamiltonian followed by the application of the correcting unitary:
\begin{eqnarray}
\Delta W & = & \mathrm{Tr}(\rho' \tilde{H} ) - \mathrm{Tr}(\rho H ) \\
& = & \mathrm{Tr}\left(\sum_i P_i \ketbra{\tilde{E_i}}{\tilde{E_i}} \sum_j \tilde{E_j} \ketbra{\tilde{E_j}}{\tilde{E_j}} \right) \nonumber\\
& & - \mathrm{Tr}\left(\sum_i P_i \ketbra{E_i}{E_i} \sum_j E_j \ketbra{E_j}{E_j} \right) \\
& = & \sum_i P_i \left(\tilde{E_i} - E_i\right),
\end{eqnarray}
and noting that $\Delta E_i = \tilde{E_i}-E_i$, 
this recovers the same work cost as if there had been no coherences at all:
\begin{equation}
\Delta W = \sum_i P_i \Delta E_i.
\end{equation}
We have compensated for our passive transformation $U$ on the Hamiltonian basis by actively applying the same transformation to the density matrix.

\section{Average work cost}
\label{app:average}
\subsection{Partial swap model of thermalisation}
\label{app:partialswap}
Consider now the process of thermalisation.
At stage $n$, when the energy levels are split by $\Delta E = n\mathcal{E}$, the associated thermal populations are given by the Gibbs distribution:
\begin{equation}
\label{eq:thermal_population}
\mathbf{P^\mathrm{th}} (n) = \dfrac{1}{1+ e^{ -\beta n \mathcal{E}}}
 \left(\begin{array}{c}
1 \\
e^{-\beta n \mathcal{E}}
\end{array}\right)
\end{equation}

Consider a stochastic transformation in which we have a probability $P_\mathrm{swap}$ of replacing the current state with the appropriate thermal state, and probability $(1-P_\mathrm{swap})$ of leaving the state alone. 
This is expressed as the stochastic matrix
\begin{equation}
\label{eq:swap_matrix}
M(n) = (1 - P_\mathrm{swap}) ~ \id + P_\mathrm{swap} M_\mathrm{th}(n),
\end{equation}
where 
\begin{equation}
\label{eq:thermal_matrix}
M_\mathrm{th}(n) = \left( \begin{array}{cc}
P_{1}^\mathrm{th}(n) & P_{1}^\mathrm{th}(n) \\
P_{2}^\mathrm{th}(n) & P_{2}^\mathrm{th}(n)
\end{array} \right)
\end{equation}
and $P_{i}^\mathrm{th}(n)$ is the $i^\mathrm{th}$ component of $\mathbf{P^\mathrm{th}}(n)$.
We verify that this matrix has the defining behaviour of a thermalising process by considering its eigenvectors, 
and noting that it evolves all probability distributions towards the thermal distribution.

Under the assumption that we raise the energy levels in such a way that the population of the levels is undisturbed, we can write a recursive relationship between the populations $\mathbf{P}{(n)}$ at the end of each stage (that is the populations having adjusted the energy level for the $n^\mathrm{th}$ time, and then allowing it to partially thermalise):
\begin{equation}
\label{eq:recursive_relation}
\mathbf{P}{(n)} =  \big(M\left(n\right)\big)^{t(n)}\ \mathbf{P}{(n-1)},
\end{equation}
where $t(n)$ is the number of times we apply the partial thermalisation matrix at stage $n$.

We attempt instead to express this energy level population as the ideal thermal distribution $\mathbf{P^{(n)}_\mathrm{th}}$ perturbed by a small difference. 
\begin{eqnarray}
\mathbf{P}(n) & = & \mathbf{P^{th}}(n) + \boldsymbol\delta (n)
%
\end{eqnarray}

We calculate the correction term $\boldsymbol\delta$ explicitly by noting that the probability of not swapping with with $\mathbf{P^{th}}(n)$ in any stage is given by $(1-P_\mathrm{swap})^{t(n)}$, and so our correction is to subtract this amount of the thermal population and add instead the same amount of the population of the previous stage:
\begin{equation}
\label{eq:trace_distance}
\boldsymbol\delta (n) = (1-P_\mathrm{swap})^{t(n)} \left(  -\mathbf{P^{th}}(n) + \mathbf{P}(n-1) \right)
\end{equation}

Further note that, as $\mathbf{P^{th}}(n)$ and $\mathbf{P}(n)$ are both valid probability distributions, the sum of the components of $\boldsymbol\delta$ must sum to zero. 
This can specifically seen to be the case here by noting that the second component of each contributing part of $\boldsymbol\delta$ is just just one minus the first component.
Indeed for a two level system, we can express $\boldsymbol\delta$ as:
\begin{equation}
\boldsymbol\delta(n) = \pm \delta(n) \left( \begin{array}{c}
1 \\
-1
\end{array} \right),
\end{equation}
where $\delta$ is a quantity known also as the variational distance (see e.g.~\cite{LevinPE09}). 
The sign of $\delta$ depends on whether we are raising or lowering the upper energy level.
Incomplete thermalisation after raising the upper energy level leaves a higher probability of being in the upper level than the true thermal probability. 
There is a lower than thermal probability when lowering the upper energy level. 

\subsection{Bounding the variational distance.}
\label{app:deltabound}
We can bound the variational distance $\delta(n)$ by noting that, since we increase the energy of the second level throughout the protocol, the effect of thermalization is always to increase the occupation probability of the first energy level $P_1^{(n-1)}\geq P_1^{(0)}$. This can be seen from Eq. \ref{eq:trace_distance}. Thus we have:
\begin{eqnarray}
\delta(n)& =& (1-p)^t \left(\dfrac{1}{1+\exp \left(-n \beta \mathcal{E} \right) } - P_1^{(n-1)} \right) \nonumber \\
& \leq & (1-p)^t \left( \dfrac{1}{1+\exp \left(-n \beta \mathcal{E} \right) } - \dfrac{1}{2} \right).
\end{eqnarray}
Having derived a bound on the variational distance we can now write the probability distribution (after some rearranging) as:
\begin{equation}
\mathbf{P}(n)  \leq  \left( \begin{array}{c}
\left(1 + (1-p)^t \right) \dfrac{1}{1 + \exp \left(-n \beta \mathcal{E} \right) } - \dfrac{1}{2}(1-p)^t \\
\left(1 + (1-p)^t \right) \dfrac{\exp \left(-n \beta \mathcal{E} \right)}{1 + \exp \left(-n \beta \mathcal{E} \right) } - \dfrac{1}{2}(1-p)^t
\end{array} \right)
\end{equation}

\subsection{Bounding the average work cost.}
\label{app:workbound}
In the worst case scenario, at each stage before raising the energy level, we are exactly $\delta$ away from the true thermal population.
Raising the energy level by $\mathcal{E}$ at stage $n$ has an associated work cost of $P_2(n)\mathcal{E} = \left(P^\mathrm{th}_2(n)+\delta\right)\mathcal{E}$.
Writing out $P^\mathrm{th}_2(n)$ explicitly (using Eq.~\ref{eq:thermal_population}), and summing over all $N$ stages in the bit reset, 
the (worst case) work cost is 
\begin{eqnarray}
&\expt{W}&=\sum_{n=0}^{N}\left(\dfrac{\exp\left( \dfrac{-n\mathcal{E}}{\kB  T}\right) }{1+\exp\left( \dfrac{-n\mathcal{E}}{\kB  T} \right) } \right. + \nonumber \\
&+&\left. (1-p)^t \left( \dfrac{1}{1+\exp \left( \dfrac{-n\mathcal{E}}{\kB T} \right) } - \dfrac{1}{2} \right) \right) \mathcal{E}.\hspace{1em}
\end{eqnarray}

For small $\mathcal{E}$, we can approximate this sum as an integral and, writing $\mathcal{E}\mathrm{d}n=\mathrm{d}E_{2}$, and $N\mathcal{E} = E_\mathrm{max}$:

\begin{eqnarray}
\expt{W} & \approx &
\int_{0}^{E_\mathrm{max}} \hspace{-1em}\mathrm{d}E_{2}
\left(\dfrac{\exp\left( -\frac{E_{2}}{\kB  T} \right) }{1+\exp\left( -\frac{E_{2}}{\kB  T}\right) }\hspace{1em}+\right. \nonumber \\*
&&+ \left. (1-p)^t \left( \dfrac{1}{1+\exp \left( \frac{-E_{2}}{\kB T} \right) } - \dfrac{1}{2} \right) \right) \nonumber \\
&=& \left(1-(1-p)^t\right) \kB T\ln\left(\dfrac{2}{1+\exp\left( -\frac{E_\mathrm{max}}{\kB  T} \right) }\right) \nonumber \\*
&&+  \dfrac{1}{2}(1-p)^t \left(E_\mathrm{max} \right)
\end{eqnarray}

Note that the term in the denominator of the argument of the logarithm is just the partition function associated with an energy level splitting of $E_\mathrm{max}$. 
If $E_\mathrm{max}\to\infty$, then this term turns to $1$, and provided we have perfect thermalisation such that if $\delta=0$, we recover the quasistatic Landauer cost $\kB T \ln 2$ (as derived in Eq.~1).

\section{Single-shot work cost}
\label{app:singleshot}
\subsection{Bounding the work cost fluctuations.}
\label{app:fluxbound}
In a single shot classical regime, we assume that at the end of each stage the system is in one of the two energy levels; and we can express each choice of energy level as a sequence of random variables $\{X_i\}_{i=1\ldots N}$.
Noting that at each stage, by raising the splitting of the energy levels by $\mathcal{E}$, if the system is in the upper energy state at a particular stage, then this stage contributes a work cost of $\mathcal{E}$.
With this in mind, it is useful to label the two energy levels as $0$ or $1$ such that the work contribution at each stage is given by $X_i\mathcal{E}$, and thus the {\em actual work cost} of a bit reset is given by the function acting on the random variables:
\begin{equation}
\label{eq:single_shot_work_X}
W(X_1, \ldots, X_N) = \mathcal{E} \sum_{i=1}^N X_i.
\end{equation}

It is possible to take the average of this function over some or even all of the random variables. 
For example, if we take the average over all $X_1\ldots X_N$ we arrive at:
\begin{eqnarray}
\expt{W(X_1, \ldots, X_N)}_{X_1 \ldots X_N}  \hspace{-5em} & & \nonumber \\ 
& = & \mathcal{E} \sum_{i=1}^N \expt{X_i}, \nonumber \\
& = & \mathcal{E} \sum_{i=1}^N P_2(i)  =  \expt{W},
\end{eqnarray}
where $\expt{W}$ is the value we would typically call the {\em average work cost}- the average work cost of the procedure calculated before we know the outcome of any of the random variables. 
This is the value that we have calculated in the prior sections of this article.

There is, in fact, a series of $N$ intermediate stages between $W$ and $\expt{W}$, in which given knowledge of the first $n$ values of $X_i$ (that is, the exact cost of the first $n$ steps of the procedure), we make an estimate of what the final work cost will be. 
This series evolves as a random walk starting at the average value $\expt{W}$ and finishing at the actual value $W$. 
Thus if the first $n$ steps of the protocol are in energy levels $X_1=x_1$, $X_2=x_2$, etc, then we write the series $D(n)$ as:
\begin{equation}
\label{eq:doob_work_function}
D(n) = \expt{W(x_1,\ldots x_n,X_{n+1}, \ldots, X_N)}_{X_{N-n}\ldots X_N}
\end{equation}

$D(n)$ undergoes a special type of random walk known as a Doob martingale~\cite{Doob40}. 
It is a martingale because at every step $n$, the expected value of the next step is the value of the current step:
\begin{equation}
\label{eq:martingale}
\expt{D(n+1)} = D(n).
\end{equation}
For Doob's martingale, this is true by construction as $W(n)$ is defined to be the expectation value of $W$ over future steps.

There is a statistical result known as the Azuma inequality~\cite{Azuma67} 
which bounds how far a martingale random walk is likely to deviate from its initial value. 
When specifically applied to a Doob martingale, this gives us the special case known as the McDiarmid inequality (see 6.10 in~\cite{McDiarmid89})
which bounds how far the actual value ($D(N) = W$) deviates from the expectation value ($D(0) = \expt{W}$):
\begin{equation}
\label{eq:McDiarmid}
P(|W - \expt{W}| \geq \omega) \hspace{0.75em} \leq  \hspace{0.75em} 2\exp\left( \dfrac{-2\,\omega^2}{\sum_{i=1}^N {|c_n|}^2} \right)
\end{equation}
where $c_n$ is the maximum amount our adjustment of the work estimate will change by knowing the outcome of the random variable $X_n$.

For pedagogical purposes we calculate this expression first for the quasistatic regime of perfect thermalisation (as discussed in the appendices of Egloff and co-authors~[13]). 
In this regime, each $X_i$ is an independent random variable, with a probability distribution given by the thermal populations associated with that energy level (Eq.~\ref{eq:thermal_population}).
Switching any particular $X_i$ from $0$ to $1$ or vice-versa will therefore have at most an effect of $\mathcal{E}$ on $W$. Hence, $c_i = \mathcal{E}$ for all $i$, and so we arrive at Eq.~7 
(repeated here for clarity):
\begin{equation*}
P(|W-\expt{W}| \geq \omega) \hspace{0.5em} \leq \hspace{0.5em} \exp\left( - \dfrac{2 \, \omega^2}{N \mathcal{E}^2} \right).
\end{equation*}

When we enter the finite time regime, in which thermalisation is only partially achieved, $X_i$ are no longer perfectly independent. 
If we treat $P_\mathrm{swap}$ as the probability over the entire period of thermalisation that we exchange our system with the thermal state, then $X_{i+1}$ will take the value of $X_{i}$ with probability $(1-P_\mathrm{swap})$, and only with probability $P_\mathrm{swap}$ will it be given by the random thermal distribution.

To calculate the impact of exchanging one stage $X_n$, we must explicitly evaluate the difference between $
D(n) = \expt{W(x_1,\ldots x_{n-1}, 0, X_{n+1}, \ldots, X_N)}_{X_{N-n}\ldots X_N}$ and $\expt{W(x_1,\ldots x_{n-1}, 1, X_{n+1}, \ldots, X_N)}_{X_{N-n}\ldots X_N}$.
To evaluate this, it is necessary to consider the expectation work cost at every stage of the protocol between $n$ and the end ($N$), and how this changes depending on the value of $X_n$.

We recall the stochastic matrix for evolution between a state $n$ and $n+1$ can be written as $M(n)$, given in Eq.~\ref{eq:swap_matrix}. 
The stochastic evolution from state $n$ to $n+k$ is thus given by the left-product of matrices
\begin{equation}
\label{eq:many_swap_matrices}
\mathcal{M}(n\to n+k) =  M(n+k-1) \ldots M(n+1) M(n).
\end{equation}

The form of $M_\mathrm{swap}(n)$ allows us to simplify this product. Over $k$ steps are there are $k+1$ possible outcome states of the system corresponding swapping with one of the thermal states $\mathbf{P}^\mathrm{th}(n+j)$ (where $j = 0\ldots k-1$) or doing nothing at all. 
Because of the nature of a swap operation (in particular because $P_\mathrm{swap}$ is independent of the state of the system), only the final swap is important- all intermediate swaps will be `overwritten'. 
This means there is always a probability of $P_\mathrm{swap}$ that the system is in the state it last had some chance of swapping with (i.e.\ $\mathbf{P}^\mathrm{th}(n+k-1)$).
Provided it has not swapped with this state, there is then a chance $P_\mathrm{swap}$ that the system has swapped into the state before it- giving an overall probability of swapping into this state of $(1-P_\mathrm{swap}) P_\mathrm{swap}$.
Working backwards with this logic, we see that the probability of the system after $k$ steps being in the state associated with swapping after $j$ steps is $(1-P_\mathrm{swap})^{k-j}P_\mathrm{swap}$.
Finally, we note that the only way for the system to have not changed at all is for it to have not swapped at any of the opportunities; and this has a probability of $(1-P_\mathrm{swap})^k$.
With all of this in mind, we can now write out $\mathcal{M}(n\to n+k)$ in a simple linear form:
\begin{eqnarray}
\mathcal{M}(n\to n+k) & = & (1-P_\mathrm{sw})^k \id \nonumber \\
&& \hspace{-5.5em} + ~ P_\mathrm{sw} \sum_{j=0}^{k-1} (1-P_\mathrm{sw})^{k-j-1} M_\mathrm{th}(n+j),
\end{eqnarray}
where $M_\mathrm{th}(i)$ (as defined in Eq.~\ref{eq:thermal_matrix}) is the matrix that perfectly exchanges any state with the thermal state $\mathbf{P}^\mathrm{th}(i)$. If $P_\mathrm{swap}=1$, then $\mathcal{M}$ reduces to the thermalising matrix for the final stage $M_\mathrm{th}(n+k-1)$.

The expected energy cost of step $n+k$ given that the state starts in the lower energy level is then given by the top right component of $\mathcal{M}$ multiplied by $\mathcal{E}$, and the expected cost if the state starts in the upper energy level is given by the lower right component multiplied by $\mathcal{E}$.
We can thus express the difference between these two values as
\begin{equation}
\mathcal{E} \left( \begin{array}{cc} 0 & 1 \end{array}\right) \mathcal{M} (n \to n+k) \left(\begin{array}{c} 1 \\ -1 \end{array} \right).
\end{equation}
Again, considering the special case $P_\mathrm{swap}=1$, we note that this value is zero for all $k\geq1$; the change in expected contribution from all future steps as a result of altering the current state is zero when the future steps are independent of the current state.

Finally, we write out the predicted difference in estimated final work cost, $c_n$, as the sum:
\begin{equation}
\label{eq:explicit_cn}
c_n = \left( \begin{array}{cc} 0 & \mathcal{E} \end{array}\right) \left(\id + \sum_{k=1}^N \mathcal{M} (n \to n+k) \right) \left(\begin{array}{c} 1 \\ -1 \end{array} \right).
\end{equation}

In this most general form, the expression is difficult to evaluate analytically.
However, noting that the components of $M_\mathrm{th}$ are the thermal populations at each stage, which are in the range $[0, 1]$, we can bound the sum:
\begin{equation}
0 \leq \sum_j (1-P_\mathrm{sw})^j P^\mathrm{th}(j) \leq \sum_j (1-P_\mathrm{sw})^j \left(\begin{array}{cc}
1 & 1 \\
1 & 1
\end{array}\right)
\end{equation}
Evaluating equation \ref{eq:explicit_cn} effectively involves subtracting the bottom right element of the matrix in the middle from the bottom left.
And as a maximal value of $|c_n|$ would be the most deleterious for our bound (causing the widest spread of work values from the mean) we therefore take the upper bound on this sum for the bottom right and the lower bound for the bottom left. 

This allows us to bound $|c_n|$ by:
\begin{equation}
\label{eq:workstepdif}
|c_n| \leq {\large\mathcal{E}} \dfrac{1-(1-P_\mathrm{sw})^{N-n}}{P_\mathrm{sw}}.
\end{equation}
This tells us that when we are far away from the end of the protocol, such that $n \ll N$, the effect of changing one stage has an effect that scales like $1/P_\mathrm{sw}$. 
Closer to the end of the protocol, the effect is diminished, as there are fewer chances to swap, which thus truncates the influence.

We consider the sum of terms $\sum_n |c_n|^2$, as required for the McDiarmid inequality (Eq.~\ref{eq:McDiarmid}):
\begin{equation}
\sum_n |c_n|^2 \leq \dfrac{\mathcal{E}^2}{P_\mathrm{sw}^2}  \sum_n \left(1-(1-P_\mathrm{sw})^{N-n}\right)^2.
\end{equation}
The leading term of the sum is $N$, and so again we can upper bound this value:
\begin{equation}
\sum_n |c_n|^2 \leq \dfrac{N \mathcal{E}^2}{P_\mathrm{sw}^2}.
\end{equation}
By taking only the first term we slightly over-estimate the importance of a change at one step. 
This approximation encodes the assumption that any change has the full number of chances to influence the state. 
This means the bound we place on the deviation away from the average value of work is not as tight as it might otherwise be.
We finally substitute this value into the McDiarmid inequality to arrive at the claim we made in Eq.~8: 
\begin{equation*}
P(|W-\expt{W}| \geq \omega) \hspace{0.5em} \leq \hspace{0.5em} \exp\left( - \dfrac{2 \,  \omega^2{P_\mathrm{sw}}^2}{N \mathcal{E}^2} \right).
\end{equation*}

\begin{figure}[tbh]
\begin{centering}
\includegraphics[width=0.45\textwidth]{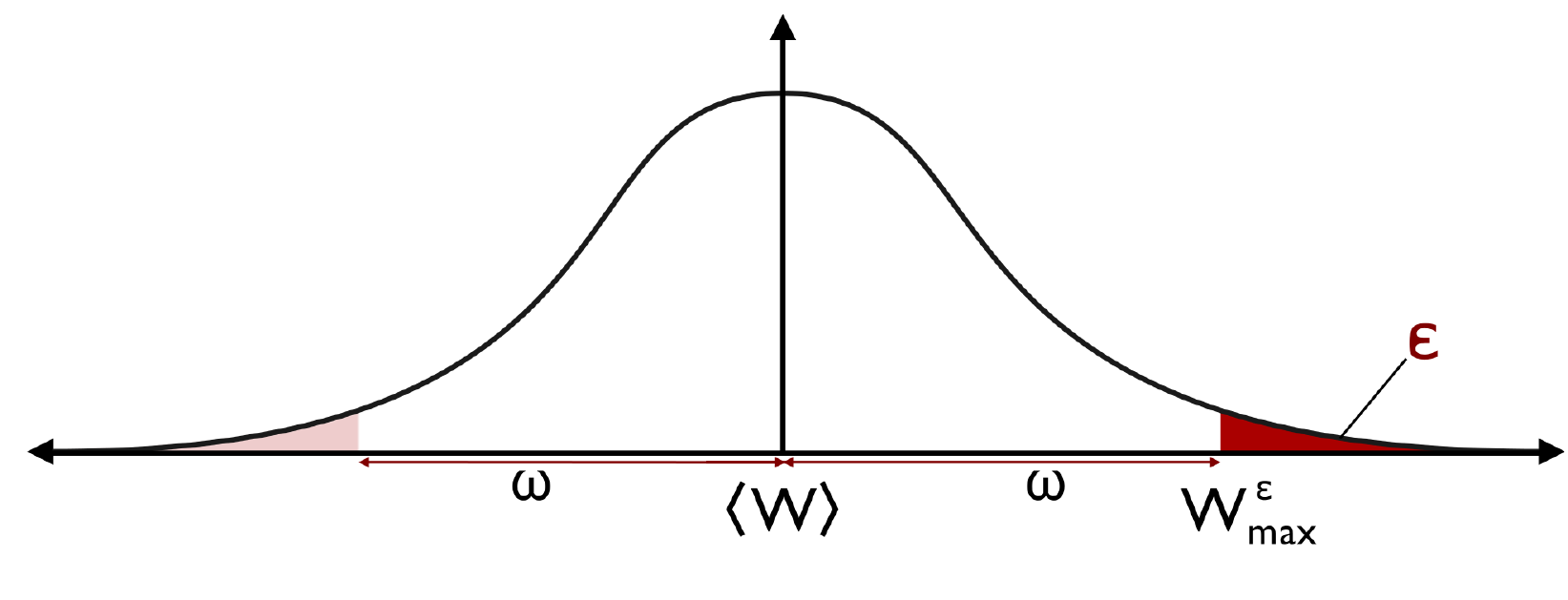}
\caption{
\label{fig:work_distribution}
{\bf An example probabilistic work distribution.} 
Work values are always below $W_\mathrm{max}^\epsilon$, except with probability of failure $\epsilon$ given by the area of the shaded region on the right. 
The sum of the areas of both the shaded regions on the left and right indicates the probability of failing to be within $\omega$ of $\expt{W}$ (as given by Eq.~8). 
}
\end{centering}
\end{figure}

\subsection{Calculating $W_\mathrm{max}^\epsilon$.}
\label{app:wemax}
We can express the above result in the language of single-shot statistics by calculating the maximum work, except with some probability of failure $\epsilon$, defined:
\begin{equation}
P(W > W_\mathrm{max}^\epsilon) := \epsilon
\end{equation}
or equivalently $W_\mathrm{max}^\epsilon$ is defined by the integral
\begin{equation}
\int_{-\infty}^{W_\mathrm{max}^\epsilon} P(W) \mathrm{d}W = 1- \epsilon.
\end{equation}

We can re-centre this definition around the expectation value of work, $\expt{W}$, such that:
\begin{equation}
P\big(W - \expt{W} > W_\mathrm{max}^\epsilon - \expt{W}\big) = \epsilon
\end{equation}
and in this form, we note that $\epsilon$ can be bounded by Eq.~8, 
with $\omega = W_\mathrm{max}^\epsilon - \expt{W}$ (see Figure~\ref{fig:work_distribution}):
\begin{equation}
\epsilon \leq \hspace{0.5em} 2\exp\left( - \dfrac{2 \,  \left(W_\mathrm{max}^\epsilon - \expt{W}\right)^2{P_\mathrm{sw}}^2}{N \mathcal{E}^2} \right).
\end{equation}
Re-arranging, we arrive at an upper bound on $W_\mathrm{max}^\epsilon$:
\begin{equation}
W_\mathrm{max}^\epsilon \leq \expt{W} + \dfrac{\mathcal{E}}{P_\mathrm{sw}} \sqrt{\dfrac{N\ln (2/\epsilon) }{2}},
\end{equation} 
or explicitly in terms of thermalising time (using Eq.~2): 
\begin{equation}
W_\mathrm{max}^\epsilon \leq \expt{W} + \dfrac{\mathcal{E}}{1-(1-p)^t} \sqrt{\dfrac{N\ln (2/\epsilon) }{2}}.
\end{equation} 
We combine this bound with the influence of finite time on $\expt{W}$ (from Eq.~5 
substituting $E_\mathrm{max} = N \mathcal{E}$) to get: 
\begin{eqnarray}
W_\mathrm{max}^\epsilon & \leq & \expt{W_\mathrm{quasi}}+ \dfrac{1}{2}(1-p)^t  N \mathcal{E}  \nonumber \\
&&+ \hspace{1em}\dfrac{1}{1-(1-p)^t} \sqrt{\dfrac{\ln (2/\epsilon) }{2N}}N\mathcal{E}. \nonumber
\end{eqnarray}

\section{Failure probability for resetting many~qubits.}
\label{app:manyqubits}
We note that our calculated $W_\mathrm{max}^\epsilon$ has failure probability upper bounded by $\epsilon$, 
such that $P(W~>~W_\mathrm{max}^\epsilon) \leq \epsilon$.
For $n$ random bits, the probability that every bit resets under this limit is given by $(1-\epsilon)^n$, 
such that we can bound:
\begin{equation}
P(W_\mathrm{tot} > nW_\mathrm{tot}^{\epsilon}) \leq 1 - (1-\epsilon)^n \leq n\epsilon,
\end{equation}
where the final inequality can be proved by induction.

In the worst case scenario, the failure of a single bit to reset under $W_\mathrm{max}^\epsilon$ causes the entire protocol to fail.
We can thus write a bound on the work cost of resetting $n$ bits as $\mathcal{W}^{\epsilon'}_\mathrm{max} \leq nW_\mathrm{max}^{\epsilon}$ where $\epsilon' = 1 - (1-\epsilon)^n$, and as $\epsilon'\leq n\epsilon$ it follows that $\mathcal{W}^{n\epsilon} \leq nW^\epsilon_\mathrm{max}$. 
Hence, when resetting $H_\mathrm{max}$ bits:
\begin{equation}
\mathcal{W}^{H_\mathrm{max}\epsilon} \leq H_\mathrm{max} W^\epsilon_\mathrm{max}.
\end{equation}


\begin{thebibliography}{10}

\bibitem{Szilard29}
L.~Szilard,
\newblock Zeitschrift f\"{u}r Physik {\bf 53}, 840 (1929).

\bibitem{Landauer61}
R.~Landauer,
\newblock IBM Journal of Research and Development {\bf 5}, 183 (1961).

\bibitem{Bennett82}
C.~H. Bennett,
\newblock International Journal of Theoretical Physics {\bf 21}, 905 (1982).

\bibitem{Frank05}
M.~P. Frank,
\newblock 2013 IEEE 43rd International Symposium on Multiple-Valued Logic {\bf
  0}, 168 (2005).

\bibitem{delRioARDV11}
L.~del Rio, J.~Aberg, R.~Renner, O.~C.~O. Dahlsten, and V.~Vedral,
\newblock Nature {\bf 474}, 61 (2011).

\bibitem{Aberg11}
J.~Aberg,
\newblock Nature Communications {\bf 4} (2011/2013),
\newblock arXiv:1110.6121.

\bibitem{DahlstenRRV11}
O.~C.~O. Dahlsten, R.~Renner, E.~Rieper, and V.~Vedral,
\newblock New Journal of Physics {\bf 13}, 053015 (2011).

\bibitem{EgloffDRV12}
D.~Egloff, O.~C.~O. Dahlsten, R.~Renner, and V.~Vedral,
\newblock (2012), arXiv:quant-ph/1207.0434.

\bibitem{Jarzynski97}
C.~Jarzynski,
\newblock Phys. Rev. Lett. {\bf 78}, 2690 (1997).

\bibitem{SekimotoS97}
K.~Sekimoto and S.~Sasa,
\newblock Journal of the Physical Society of Japan {\bf 66}, 3326 (1997).

\bibitem{SerreliLKL07}
V.~Serreli, C.~Lee, E.~R. Kay, and D.~A. Leigh,
\newblock Nature {\bf 445}, 523 (2007).

\bibitem{ThornSLS08}
J.~Thorn, E.~Schoene, T.~Li, and D.~Steck,
\newblock Physical Review Letters {\bf 100} (2008).

\bibitem{Toyabe2010}
S.~Toyabe, T.~Sagawa, M.~Ueda, E.~Muneyuki, and M.~Sano,
\newblock Nature Physics {\bf 6}, 988 (2010).

\bibitem{AurellMM11}
E.~Aurell, C.~Mejia-Monasterio, and P.~Muratore-Ginanneschi,
\newblock Physical Review Letters {\bf 106} (2011),
\newblock {arXiv}: 1012.2037.

\bibitem{BerutAPCDL12}
A.~B\'{e}rut {\em et~al.},
\newblock Nature {\bf 483}, 187 (2012).

\bibitem{BergliGK13}
J.~Bergli, Y.~M. Galperin, and N.~B. Kopnin,
\newblock Physical Review E {\bf 88} (2013),
\newblock {arXiv}: 1306.2742.

\bibitem{Zulkowski13}
P.~R. Zulkowski and M.~R. DeWeese,
\newblock {arXiv}:1310.4167 [cond-mat]  (2013),
\newblock {arXiv}: 1310.4167.

\bibitem{Koski14}
J.~V. Koski, V.~F. Maisi, J.~P. Pekola, and D.~V. Avering,
\newblock (2014), 1402.5907.

\bibitem{AlickiHHH04}
R.~Alicki, M.~Horodecki, P.~Horodecki, and R.~Horodecki,
\newblock Open Systems \& Information Dynamics {\bf 11}, 205 (2004).

\bibitem{Kieu04}
T.~D. Kieu,
\newblock Phys. Rev. Lett. {\bf 93} (2004).

\bibitem{quan08}
H.~T. Quan and H.~Dong,
\newblock {arXiv}:0812.4955 [cond-mat]  (2008),
\newblock {arXiv}: 0812.4955.

\bibitem{Berry09}
M.~V. Berry,
\newblock Journal of Physics A: Mathematical and Theoretical {\bf 42}, 365303
  (2009).

\bibitem{Deffner14}
S.~Deffner, C.~Jarzynski, and A.~del Campo,
\newblock Physical Review X {\bf 4} (2014),
\newblock {arXiv}: 1401.1184.

\bibitem{DelCampo13}
A.~del Campo, J.~Goold, and M.~Paternostro,
\newblock {arXiv}:1305.3223 [cond-mat, physics:quant-ph]  (2013),
\newblock {arXiv}: 1305.3223.

\bibitem{ZimanSBHSG01}
M.~Ziman {\em et~al.},
\newblock (2001), arXiv:quant-ph/0110164.

\bibitem{ScaraniZSGB02}
V.~Scarani, M.~Ziman, P.~\v{S}telmachovi\v{c}, N.~Gisin, and V.~Bu\v{z}ek,
\newblock Physical Review Letters {\bf 88}, 097905 (2002), 0110088.

\bibitem{LevinPE09}
D.~A. Levin, Y.~Peres, and E.~L. Wilmer,
\newblock {\em {Markov chains and mixing times}} (American Mathematical Soc.,
  2009).

\bibitem{McDiarmid89}
C.~McDiarmid,
\newblock London Mathematical Society Lecture Note Series {\bf 141} (1989).

\bibitem{FaistDOR12}
P.~{Faist}, F.~{Dupuis}, J.~{Oppenheim}, and R.~{Renner},
\newblock ArXiv e-prints  (2012), 1211.1037,
\newblock arXiv:1211.1037.

\end{thebibliography}
\end{document}